\documentclass[aps,prl,twocolumn,showpacs,floatfix,superscriptaddress]{revtex4}
\usepackage{graphicx}

\newcommand {\be} {\begin{equation}} 
\newcommand {\ee} {\end{equation}} 
\newcommand {\Be}{\begin{eqnarray*}}
\newcommand {\Ee} {\end{eqnarray*}}
\newcommand {\bey} {\begin{eqnarray}} 
\newcommand {\eey} {\end{eqnarray}} 

\begin{document}

\title{On the universality of anomalous one-dimensional heat conductivity}

\author{Stefano Lepri}
\email{stefano.lepri@unifi.it}
\affiliation{Istituto Nazionale di Fisica della Materia-UdR
Firenze, via G. Sansone 1 I-50019 Sesto Fiorentino, Italy}

\author{Roberto Livi}
\affiliation{Dipartimento di Fisica, via G. Sansone 1 I-50019, Sesto
Fiorentino, Italy }
\affiliation{Istituto Nazionale di Fisica della Materia-UdR
Firenze, via G. Sansone 1 I-50019 Sesto Fiorentino, Italy}

\author{Antonio Politi}
\affiliation{Istituto Nazionale di Ottica Applicata, largo E. Fermi 6
I-50125 Firenze, Italy}
\affiliation{Istituto Nazionale di Fisica della Materia-UdR
Firenze, via G. Sansone 1 I-50019 Sesto Fiorentino, Italy}

\date{\today}

\begin{abstract}

In one and two dimensions, transport coefficients may diverge in the 
thermodynamic limit due to long--time correlation of the corresponding
currents. The effective asymptotic behaviour is addressed with reference to the
problem of heat transport in $1d$ crystals, modeled by chains of classical
nonlinear oscillators. Extensive accurate equilibrium and nonequilibrium 
numerical simulations confirm that the finite-size thermal conductivity
diverges with the system size $L$ as $\kappa \propto L^\alpha$. However, the
exponent $\alpha$ deviates systematically from the theoretical prediction
$\alpha=1/3$  proposed in a recent paper [O. Narayan, S. Ramaswamy,
Phys. Rev. Lett. {\bf 89}, 200601 (2002)]. 
\end{abstract}


\maketitle  

Strong spatial constraints can significantly alter transport properties. The
ultimate reason is that the response to external forces depends on statistical
fluctuations which, in turn, crucially depend on the system dimensionality $d$.
A relevant example is the anomalous behaviour of heat conductivity for $d\le
2$. After the publication of the first convincing numerical evidence of a
diverging thermal conductivity in anharmonic chains \cite{LLP97}, this issue
attracted a renovated interest within the theoretical community. A fairly
complete overview is given in Ref.~\cite{LLP02}, where the effects of
lattice dimensionality on the breakdown of Fourier's law are discussed as well.  
Anomalous behaviour means both a nonintegrable algebraic decay of equilibrium
correlations of the heat current $J(t)$ (the Green-Kubo integrand) at large 
times $t\to \infty$ and a divergence of the finite-size conductivity 
$\kappa(L)$ in the $L\to \infty$ limit. This is very much reminiscent of the
problem of long-time tails in fluids \cite{PR75} where, in low spatial
dimension, transport coefficients may {\it not exist at all}, thus implying a
breakdown of the phenomenological constitutive laws of hydrodynamics. 
In $1d$ one finds
\be 
\kappa(L) \propto L^\alpha   \quad, \qquad
\langle J(t)J(0)\rangle \propto t^{-(1+\delta)}  \quad ,  
\label{anomal}
\ee
where $\alpha >0$, $-1<\delta < 0$, and  $\langle \,\, \rangle$ is the
equilibrium average. For small applied gradients, linear-response theory allows
establishing a connection between the two
exponents. By assuming that $\kappa(L)$ can be estimated by cutting-off the
integral in the Green-Kubo formula at the ``transit time'' $L/v$ ($v$ being
some propagation velocity of excitations), one obtains $\kappa \propto
L^{-\delta}$ i.e. $\alpha=-\delta$. 

Determining the asymptotic dependence of heat conductivity is not only
important for assessing the universality of this phenomenon, but may be also
relevant for predicting transport properties of real materials. For instance,
recent molecular dynamics results obtained with phenomenological carbon
potentials indicate an unusually high conductivity of single-walled
nanotubes \cite{BKT00}: a power-law divergence with the tube length has been
observed with an exponent very close to the one obtained in simple $1d$
models \cite{M02}. 

The analysis of several models \cite{LLP02} clarified that anomalous
conductivity should occur generically whenever momentum is conserved. For
lattice models, this amounts to requiring that at least one acoustic phonon
branch is present in the harmonic limit. The only known exception is the
coupled-rotor model, where normal transport \cite{GLPV00} is believed to
arise as a consequence of the boundedness of the potential.

It is thus natural to argue about the universality of the exponent $\alpha$.
On the one hand, there exist two theoretical predictions, namely $\alpha=2/5$,
which follows from self-consistent mode-coupling theory~\cite{E91,LLP98}, and
$\alpha=1/3$, obtained by Narayan and Ramaswamy~\cite{NR02} by performing a
renormalization group calculation on the stochastic hydrodynamic equations for
a 1$d$ fluid. On the other hand, the available numerical data for $\alpha$
range from 0.25 to 0.44. The most convincing confirmation of the 1/3-value
have been obtained by simulating a one-dimensional gas of hard-point particles
with alternating masses~\cite{H99,GNY02} and random-collision
models~\cite{DN03}. In the former case, a careful determination of the
scaling exponent is however hindered by the presence of large finite-size
corrections that are still sizeable for ${\cal O}(10^4)$ particles. As a matter
of fact, other authors~\cite{D01} report significantly smaller estimates
($\alpha \simeq 0.25$). This anomaly is possibly due to the lack of
microscopic chaos in that model~\cite{GNY02}. The results obtained for models 
of $1d$ crystals are more controversial, but consistently larger than 1/3. For
instance, in the case of the Fermi-Pasta-Ulam (FPU) chain, the best estimate 
sofar is $\alpha\simeq 0.37$ \cite{LLP98,GY02}.

However, with the only exception of Ref.~\cite{GNY02}, all numerical
investigations limit themselves to fitting the scaling behaviour
in a suitable range, without determining possible finite-size corrections,
so that none of them can be fully trusted. In view of the general relevance
of establishing the existence of one or more universality classes, in the
present paper we present far more accurate simulations which allow determining
the effective exponents for different lengths and frequencies. We anticipate
that $\alpha$ is definitely larger than 1/3 in a 1d crystal model and possibly
in agreement with the mode-coupling prediction.

We consider an array of $N$ point-like identical atoms ordered along a line. 
The position of the $n$-th atom is denoted with $x_n$, while its mass is fixed,
without loss of generality, equal to unity. By further assuming that
interactions are restricted to nearest-neighbour pairs, the equations of motion
write
\begin{equation} 
{\ddot x}_n = - F_n + F_{n-1} \quad ; \qquad 
F_n=- V'(x_{n+1} - x_n)  \, ,
\label{eqmot} 
\end{equation}  
where $V'(z)$ is a shorthand notation for the first derivative of the
the interparticle potential $V$ with respect to $z$. The microscopic expression
of the heat current is
\begin{equation}
J = \sum_n \left[\frac{1}{2} (x_{n+1} - x_n) ({\dot x}_{n+1} + {\dot x}_n) \,
     F_n + {\dot x}_n h_n \right] \, .
\label{hf2}
\end{equation}
where $h_n$ is a suitably defined local energy \cite{LLP02}. For small
oscillations (compared to the lattice spacing $b =L/N$), the
second term can be neglected and $x_n-x_{n-1}\simeq b$,
so that Eq.~(\ref{hf2}) can be approximated by
\begin{equation}
J \;\simeq\;  {b\over 2}  \sum_n ({\dot x}_{n+1} + {\dot x}_n) \, F_n \quad .
\label{jsolid}
\end{equation}

The customary way to evaluate the thermal conductivity $\kappa$ is through 
the Green-Kubo formula  
\be
\kappa_{GK}\,=\,\frac{1}{k_BT^2 }\lim_{t \to \infty}
\int_0^t d\tau \, \lim_{L\to \infty} L^{-1}\langle J(\tau)J(0)\rangle \quad.  
\ee  
A crucial, sometimes overlooked \cite{PC00}, point is that such formulae
are formally identical for different statistical ensembles, but the 
definition of $J$ differs, because of ``systematic" contributions associated
with other conservation laws that must be subtracted out \cite{G60}.  
For instance, expression (\ref{jsolid}) is correct in the microcanonical ensemble
with zero total momentum, while in the canonical ensemble (for large $N$) it is
\be
J \; =\; 
{b\over 2} \sum_n ({\dot x}_{n+1} + {\dot x}_n) \, F_n
\,-\, bv_0 \langle \sum_n F_n  \rangle \quad ,
\ee
$v_0$ being the center--of--mass velocity. This choice ensures 
that the autocorrelation of $J$ vanishes for $t\to \infty$.

With reference to Eq.~(\ref{anomal}), the possibly anomalous behavior can be
analyzed by computing the power spectrum of the heat current $J$.
Since we are interested in the long--wavelength and small--frequency behavior,
it is convenient to consider a highly nonlinear model in the hope that the
asymptotic regime sets in over shorter time and space scales. Moreover, it is
advisable to work with a computationally simple expression of the force.
The best compromise we have found is the quartic Fermi-Pasta-Ulam potential
\begin{equation}
V(z) \;=\;  {1\over 4}\, (z-a)^4 \quad .
\label{fpu4}
\end{equation}
Indeed, after the change of coordinates $x_n=u_n+na$ \cite{note1}, the physical
distance $a$ disappears from the equations of motion for $u_n$. This 
model has no free parameters: since the potential expression is homogeneous,
the dynamics is invariant under coordinate rescaling, so that the energy per
particle $e$ can be set, without loss of generality, equal to 1.

\begin{figure}[ht!]
\includegraphics[clip,width=6.5cm]{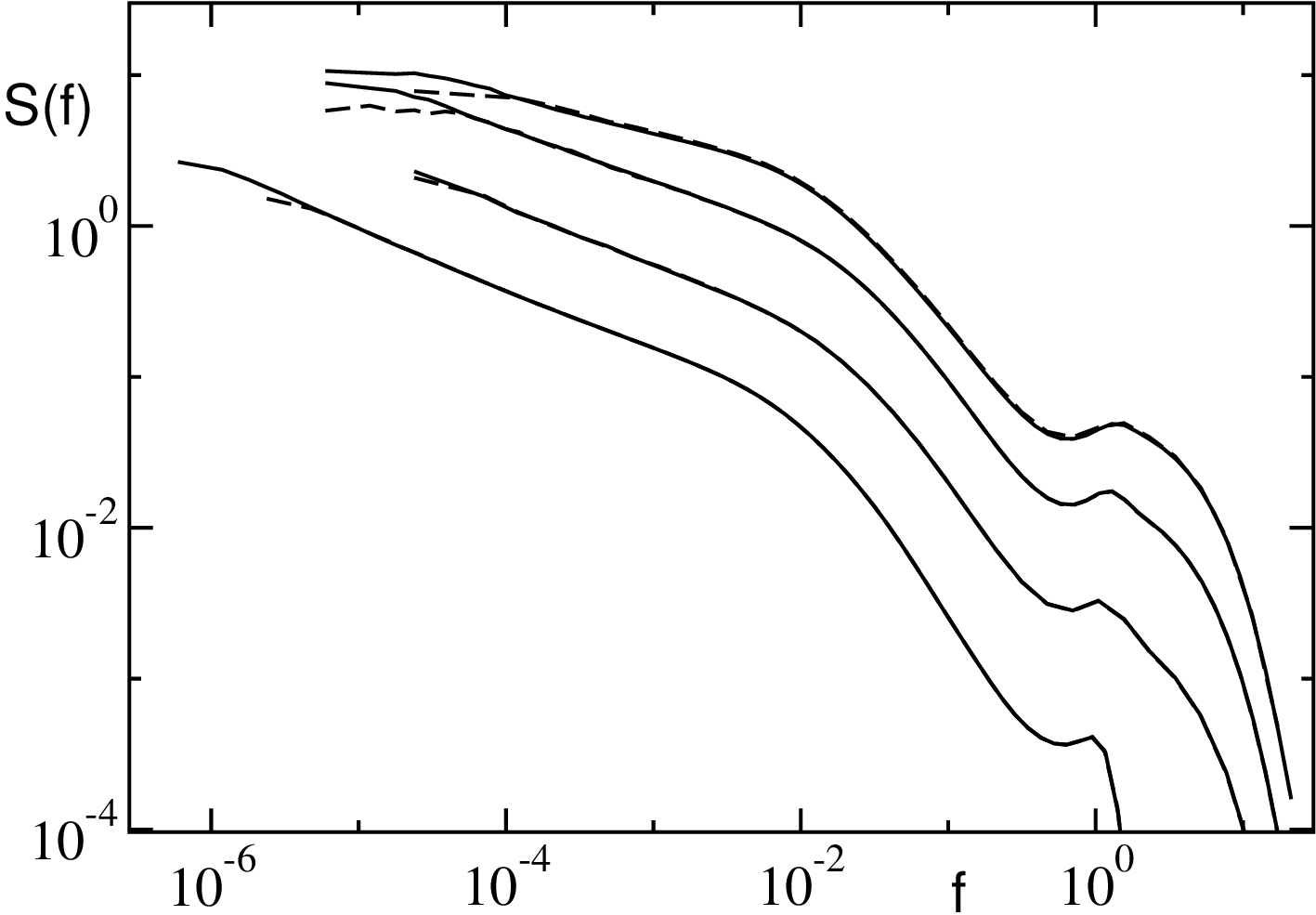}
\caption{
Power spectra of the flux $J$ as defined in Eq.~(\ref{jsolid}). The lowermost
curves refer to the purely quartic FPU model (\ref{fpu4}) with $N=2048$ (solid)
and 1024 (dashed). The upper curves correspond to the repulsive FPU model --
Eq.~(\ref{repfpu}) -- for $a=2$, 2.3, and 2.5 (from top to bottom)   
and $N=1024$ (solid) or 512 (dashed). All microcanonical simulations are
performed for the same energy density $e=1$ with time step $h=0.05$
for $10^6-10^7$ steps. For clarity, the curves have been
arbitrarily shifted along the vertical axis.} 
\label{fig1}
\end{figure}

First, we have performed equilibrium microcanonical simulations by integrating
Eqs.~(\ref{eqmot}) (with periodic boundary conditions) 
with a fourth--order symplectic algorithm \cite{MA92}.
The power spectra $S(f)$ of $J$ are reported in Fig.~\ref{fig1}. The lowest curves  are data
for the quartic FPU model obtained by averaging over 30,000 random initial
conditions. In order to further decrease statistical fluctuations, a binning of
the data over contiguous frequency intervals has been performed as well.  

The long--time tail (\ref{anomal}) manifests itself as a power--law divergence
$f^\delta$ in the low-$f$ region. By comparing the results obtained for
different numbers of particles, one can clearly see that finite-size
corrections are negligible above a size-dependent frequency $f_c(N)$. 
By fitting the data in the scaling range $[f_c(N),f_s]$, where $f_s \simeq
10^{-3}$ , we find $\delta=-0.39(6)$. These values are consistent with
previous, less-accurate, findings for similar models, such as the standard FPU
\cite{LLP98,GY02} and the diatomic Toda \cite{H99} chains, thus confirming the
expectation that they all belong to the same universality class.

\begin{figure}[ht!]
\includegraphics[clip,width=6.5cm]{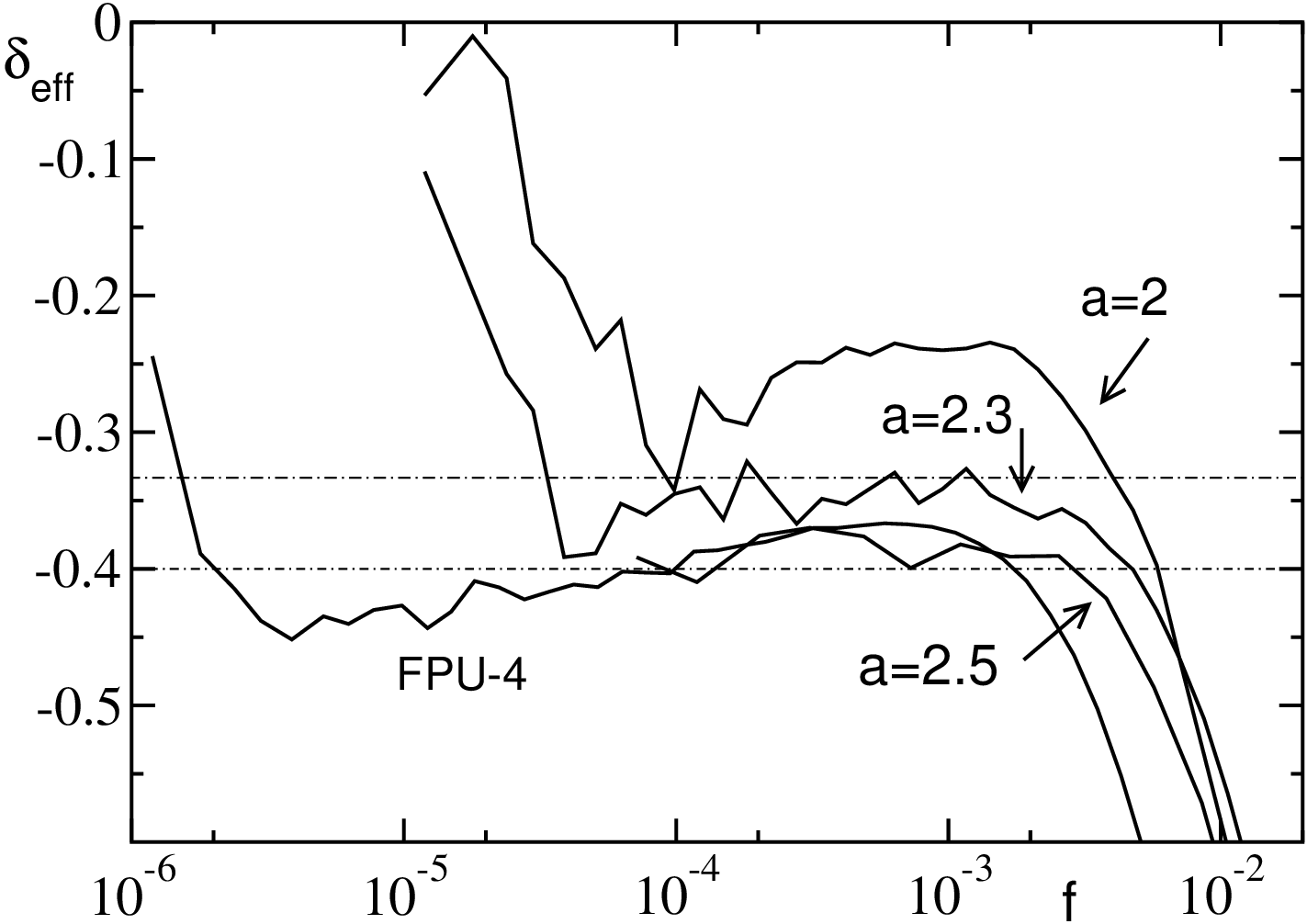}
\caption{
The logarithmic derivative $\delta_{\rm eff}$ of the energy-flux spectrum
versus the frequency $f$ for the pure FPU quartic potential -- FPU-4  with
$N=2048$ -- and for model (\ref{repfpu}) with $a=2$, $a=2.3$, and $a=2.5$  and
$N=1024$, respectively. The two horizontal lines correspond to the theoretical
predictions $-1/3$ and $-2/5$. The statistical error is on the order of the 
observed irregular fluctuations.} 
\label{fig:deriv}
\end{figure}
 
In order to perform a more stringent test of the scaling behavior, we have
determined the logarithmic derivative
\begin{equation}
\delta_{\rm eff}(f) \;=\; \frac{d \ln S}{d \ln f} \quad.
\label{deff}
\end{equation}
for different frequencies. Since finite-size effects are responsible for the
saturation of $S(f)$ when $f \to 0$, $f_c(N)$ can be identified (see
Fig.~\ref{fig:deriv}) as the frequency below which $\delta_{\rm eff}$ starts
growing towards zero. Above $f_c$, the quality of our numerical data allows
revealing a slow but systematic decrease of $\delta_{\rm eff}$ upon decreasing
$f$, which approaches $-0.44$, a value that is incompatible not only with the
renormalization-group prediction of Ref.~\cite{NR02}, but also with the result
of mode--coupling theory~\cite{E91,LLP98}. Furthermore, convergence seems not
fully achieved in the accessible frequency range.

Accordingly, it is advisable to look at thermal conductivity by means of
nonequilibrium simulations too. It is sufficient to measure the heat flux in a
system put in contact with two heat reservoirs operating at different
temperatures $T_+$ and $T_-$. Several methods have been proposed based on both
deterministic and stochastic algorithms \cite{LLP02}. Regardless of the actual
thermostatting scheme, after a  transient, an off-equilibrium stationary state
sets in, with a net heat current flowing through the lattice. The finite-size
thermal conductivity $\kappa (L)$ is then estimated as the ratio between the
average flux $\overline J$ and the overall temperature gradient $(T_+-T_-)/L$.
Notice that, by this definition, $\kappa$ amounts to an effective transport
coefficient including both boundary and bulk scattering mechanisms. 

\begin{figure}[ht!]
\includegraphics[clip,width=6.5cm]{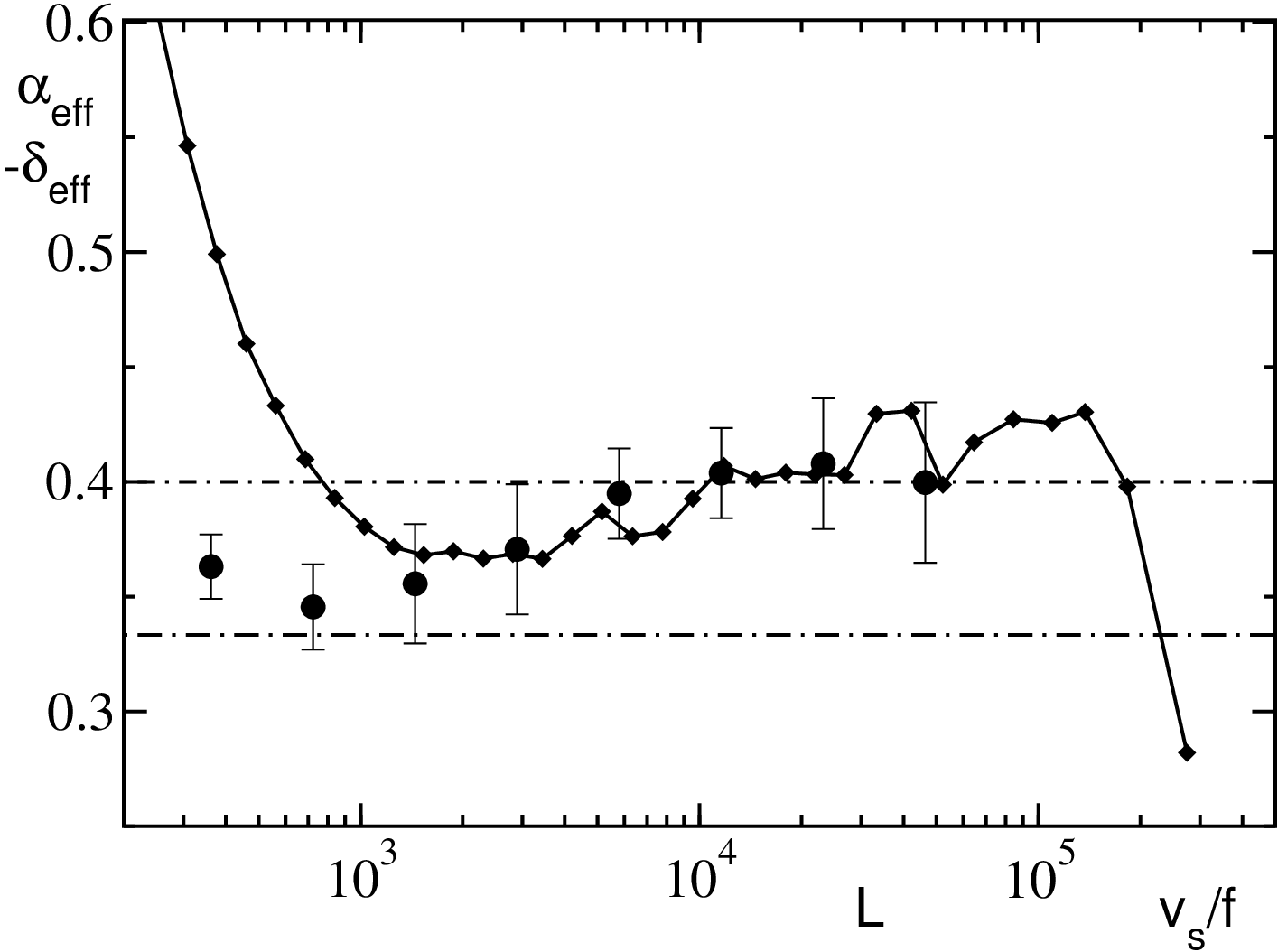}
\caption{
Quartic FPU model: the effective exponent $\alpha_{\rm eff}$ of the finite-size
conductivity for $T_+=1.2, T_-=0.8$ (full dots), compared with the results
($-\delta_{\rm eff}$) of equilibrium simulations. The two horizontal lines
correspond to the theoretical predictions, 1/3 and 2/5.} 
\label{fig:scaling}
\end{figure}

We have used the Nos\'e--Hoover thermostats described in detail in
Ref.~\cite{LLP02}. In order to fasten the convergence towards the stationary
state, the initial conditions have been generated by thermostatting each
particle to yield a linear temperature profile (see \cite{LLP97}). This method
is very efficient, especially in long chains, when bulk thermalization may be
significantly slow. The heat flux $\overline J$ has been obtained by
integrating the equations over more than 10$^6$ time units and by further
averaging over a set of about 30 initial conditions. Simulations of the
quartic FPU model with chains of length up to $65536$ sites and free boundary
conditions exhibit again a systematic increase of the effective exponent
\be 
\alpha_{\rm eff}(L) \;=\; \frac{d \ln \kappa}{d \ln L} , 
\label{aeff}
\ee
as it can be seen from the full dots in Fig.~\ref{fig:scaling}, although
one can also observe that the four rightmost values are in very good agreement 
with the mode--coupling exponent.

In order to compare more closely equilibrium and nonequilibrium simulations, 
one can assume, following the argument exposed below Eq.~(\ref{anomal}), 
that the finite--size conductivity $\kappa(L)$ is determined by correlations up
to time $\tau = L/v_s$, where $v_s$ is the sound velocity. This means that the
frequency $f$ can be turned into a length $L=v_s/f$. It might be argued that 
the absence of a quadratic term in (\ref{fpu4}) prevents a straightforward 
definition of such a velocity in the $T=0$ limit; nevertheless, it has been 
shown \cite{AC03} that an effective phonon dispersion relation at finite energy
density can be evaluated for model (\ref{fpu4}), yielding $v_s = 1.308$ at 
$e=1$. Using this value,
we can ascertain that, at least for $N>1000$, there is an excellent agreement
between the two approaches (see again Fig.~\ref{fig:scaling}).

The data presented so far rule out the value predicted in Ref.~\cite{NR02} for 
the model (\ref{fpu4}). On the other hand, such prediction is consistent with
numerical results for hard-core interaction \cite{GNY02,DN03}. In order to test
for universality we thus tried to bridge the two classes of models by
introducing a strong repulsive potential. This has also the merit to 
remove one of the main drawbacks of a potential like (\ref{fpu4}), namely that
negative values of $x_{n+1}-x_{n}$ are allowed, i.e. that particles can
formally cross each other if $x_n$ is interpreted as their actual position. 
This unphysical feature may somehow be circumvented
by introducing explicitly a physical distance. For this purpose, we have added
to the FPU potential a repulsive term of the Lennard-Jones (LJ) form
\begin{equation}
V(z) \;=\;  {1\over 4}\, (z-a)^4 + \frac{1}{12}{1\over z^{12}}  - V_0,
\label{repfpu}
\end{equation}
where $V_0$ is a suitable constant needed to set the minimum of the energy
equal to 0. The core repulsion introduces a further time-scale, namely that of
mutual ``collisions" between particles \cite{note2}.
Upon increasing $a$, at fixed energy, the role of the repulsive term becomes
negligible and model (\ref{repfpu}) reduces to the purely quartic FPU 
(\ref{fpu4}). For instance, in the region where $V(z)<e$ the LJ energy 
contribution can be as large as 0.57 for $a=2$ and $e=1$, but it is at most 
0.028, when $a$ is increased
to 2.5. Upon decreasing $a$, the LJ term progressively affects the
high-frequency spectral range. This is because core repulsion becomes
more relevant close to the minimum. 

In this context, one should, in principle, refer to the general heat-flux
expression (\ref{hf2}) that, in the limit of pure hard--points reduces to 
$\sum_n
\dot x_n^3/2$. Nevertheless, in the parameter range investigated hereby, the
spectra of this quantity never exceed 10\%  of the spectra of  $(\ref{jsolid})$
in all frequency channels and, more importantly,   hardly show any singular
low-frequency behaviour. We have therefore kept determining the power spectrum 
of the flux as defined in Eq.~(\ref{jsolid}).

The effect of the LJ term on the low-frequency behavior of $S(f)$ can be 
appreciated already for $a =2.3$. A direct fitting of the 3 upmost curves in
Fig.~\ref{fig1} (in the available scaling ranges) yields $\delta$ decreasing
from $-0.25(0)$ for $a=2.0$ up to $-0.37(8)$ for $a=2.5$. 
Having averaged the spectra over more than 5000 different samples, it is 
possible to investigate the convergence of each $\delta$-value through
the effective exponent (\ref{deff}). Like in the quartic FPU model, one can see
from Fig.~\ref{fig:deriv} that $\delta_{\rm eff}$ decreases upon decreasing
frequency, even on the smaller available range. Although on the basis of 
numerics alone one cannot exclude that the asymptotic $\delta$-value depends 
on $a$, it is wiser to conjecture that the stronger the LJ term,
the slower is the onset of the asymptotic regime.

Altogether our simulations do not confirm the claim contained in 
Ref.~\cite{NR02} that the hydrodynamic theory accounts for all $1d$ models. 
The exponent $\alpha$ is found to be definitely larger than the expected 
value 1/3 and certainly closer to the mode--coupling estimate 2/5. Anyhow, 
the systematic deviations shown in Fig.~\ref{fig:deriv} make also the 
convergence to this latter value somehow questionable. In addition, we have 
also shown that the low-frequency power--law behavior is strongly influenced 
by the presence of a hard--core repulsion term: even small variations of the 
spatial scale associated with the equilibrium distance between interacting 
oscillators enhance finite size effects and slow down convergence with respect 
to the purely anharmonic model.
This scenario rather suggests that at least two different universality classes
may exist, although their physical origin is up to know unclear.  

A similar analysis should now be applied to the other models recently
considered, in order to determine how much of the observed mutual fluctuations
are the result of finite-size corrections. However, since we have basically
reached the limit of our computing facilities letting a cluster of 48 PCs run
for two months, it is also clear that more refined analytic
estimates have to be worked out to shed light on this puzzling scenario.

This work is partially supported by the INFM-PAIS project {\it Transport
phenomena in low-dimensional structures} and by the EU network LOCNET,
Contract No. HPRN-CT-1999-00163.


\begin{thebibliography}{00}

\bibitem{LLP97} S. Lepri, R. Livi, A. Politi, Phys. Rev. Lett.
{\bf 78}, 1896 (1997).

\bibitem{LLP02} S. Lepri, R. Livi, A. Politi, Phys. Rep. {\bf 377}, 1 (2003).

\bibitem{PR75} Y. Pomeau, R. R\'esibois, Phys. Rep.  {\bf 63}, 19 (1975) .

\bibitem{BKT00} S. Berber, Y. Kwon, D. Tomanek, Phys. Rev. Lett.
{\bf 84}, 4613 (2000).

\bibitem{M02} S. Maruyama, Physica B {\bf 323}, 193 (2002).

\bibitem{GLPV00} C. Giardin\`a {\it et al.}
Phys. Rev. Lett. {\bf 84 },  2144 (2000);
O. V. Gendelman, A. V. Savin, 
{\it ibid.} {\bf 84 },  2381 (2000).

\bibitem{E91} M.H. Ernst, Physica D {\bf 47}, 198 (1991).

\bibitem{LLP98} S. Lepri, R. Livi, A. Politi, Europhys. Lett. {\bf 43},
271 (1998). 

\bibitem{NR02}  O. Narayan, S. Ramaswamy, Phys. Rev. Lett.
{\bf 89},  200601 (2002).

\bibitem{H99} T. Hatano, Phys. Rev. E {\bf 59},  R1 (1999).

\bibitem{GNY02} P. Grassberger, W. Nadler, L. Yang, 
Phys. Rev. Lett. {\bf 89}, 180601 (2002). 

\bibitem{DN03}  J.M. Deutsch, O. Narayan, Phys. Rev. E {\bf 68}, 
010201 (2003). 

\bibitem{D01} A. Dhar, Phys. Rev. Lett. {\bf 86}, 3554 (2001);
G. Casati, T. Prosen, Phys. Rev. E {\bf 67} 015203(R) (2003)

\bibitem{GY02} P. Grassberger, L. Yang, cond-mat/0204247.

\bibitem{PC00} T. Prosen and D.K. Campbell,  Phys. Rev. Lett. {\bf 84},
2857 (2000).

\bibitem{G60} M. S. Green, Phys. Rev. {\bf 119},  829 (1960).

\bibitem{note1} The average lattice spacing $b$ does not need to coincide with
the minimum $a$ of $V$. This happens only if no pressure is
exerted. As we do not expect pressure to play any role on the
scaling behaviour of physical observables, for simplicity, we always
fix $b=a$.

\bibitem{MA92} R.I. Mclachlan, P. Atela, Nonlinearity {\bf 5},  541 (1992).

\bibitem{AC03} C. Alabiso, M. Casartelli, J. Phys. A: Math. Gen.
{\bf 34} 1223 (2001).

\bibitem{note2} 
In the absence of hard--core the very definition of collision is somehow
arbitrary. We say that a collision occurs when $x_{n+1}-x_n$ is
equal to the distance at which the LJ force term is 10 times larger than the
FPU one. 

\end{thebibliography}
\end{document}